\documentclass{aa}  

\usepackage{txfonts}
\usepackage[colorlinks=true, citecolor=blue]{hyperref}
\usepackage{epsf,palatino,url,wrapfig,array,setspace,amsmath,amssymb,fancyhdr,multirow,lscape,appendix,rotating}
\usepackage{graphics,epsfig,txfonts}
\usepackage{graphicx}

\usepackage{longtable}
\usepackage{amssymb}
\usepackage{xcolor}
\usepackage{color}
\usepackage{lipsum}
\usepackage{textcomp}
\usepackage{threeparttable,booktabs}
\usepackage{enumerate}
\usepackage{placeins}
\usepackage{float}
\usepackage[normalem]{ulem}
\usepackage{soul}
\usepackage{blindtext}

\usepackage{tikz,hyperref}
\usepackage{academicons}

 \newcommand{\swift}{{\it Swift}\xspace}

\newcommand{\fermi}{{\it Fermi}\xspace}

\newcommand{\txs}{TXS~0506+056 \xspace}

\begin{document}

   \title{TeV pion bumps in the gamma-ray spectra of flaring blazars}

   \author{M. Petropoulou
          \inst{1,2}
          \and
          A. Mastichiadis\inst{1} 
        \and
          G. Vasilopoulos\inst{1,2}
        \and 
        D. Paneque\inst{3}
        \and  
         J.~Becerra~González\inst{4,5}
        \and 
          F.~Zanias\inst{1}   
      }

    \institute{Department of Physics, National and Kapodistrian University of Athens, University Campus Zografos, GR 15784, Athens, Greece  \\
    \email{mpetropo@phys.uoa.gr}   \\ 
  \and 
   Institute of Accelerating Systems \& Applications, University Campus Zografos, Athens, Greece \\  \and 
    Max-Planck-Institut für Physik, 80805 München, Germany \\ 
  \email{dpaneque@mppmu.mpg.de} \\ \and
  Instituto de Astrof\'isica de Canarias (IAC), E-38200 La Laguna, Tenerife, Spain \\
  \email{jbecerra@iac.es}
  \and  Universidad de La Laguna (ULL), Departamento de Astrof\'isica, E-38206 La Laguna, Tenerife, Spain \\  
             }

   \date{Received XXX; accepted XXX}

 
  \abstract
   {
Very high-energy (VHE, $E>100$ GeV) observations of the blazar Mrk 501 with MAGIC in 2014 provided evidence for an unusual narrow spectral feature at about 3 TeV during an extreme X-ray flaring activity. The one-zone synchrotron-self Compton scenario, widely used in blazar broadband spectral modeling, fails to explain the narrow TeV component.
   }
   {
Motivated by this rare observation, we propose an alternative model for the production of narrow features in the VHE spectra of flaring blazars. These spectral features may result from the decay of neutral pions ($\pi^0$ bumps) that are in turn produced via interactions of protons (of tens of TeV energy) with energetic photons, whose density increases during hard X-ray flares. 
   }
   {
We explored the conditions needed for the emergence of narrow $\pi^0$ bumps in VHE blazar spectra during \mbox{X-ray} flares reaching synchrotron energies $\sim 100$~keV using time-dependent radiative transfer calculations. We focused on high-synchrotron peaked (HSP) blazars, which comprise the majority of VHE-detected extragalactic sources. 
   }
   {
We find that synchrotron-dominated flares with peak energies $\gtrsim100$~keV can be ideal periods for the search of $\pi^0$ bumps in the VHE spectra of HSP blazars. The flaring region is optically thin to photopion production, its energy content is dominated by the relativistic proton population, and the inferred jet power is highly super-Eddington. Application of the model to the spectral energy distribution of Mrk 501 on MJD 56857.98 shows that the VHE spectrum of the flare is described well by the sum of a synchrotron self-Compton (SSC) component and a distinct $\pi^0$ bump centered at 3 TeV. Spectral fitting of simulated SSC+$\pi^0$ spectra for the Cherenkov Telescope Array  (CTA) show that a $\pi^0$ bump could be detected at a 5$\sigma$ significance level with a 30-min exposure.
  }
  {A harder VHE $\gamma$-ray spectrum than the usual SSC prediction or, more occasionally, a distinct narrow bump at VHE energies during hard X-ray flares, can be suggestive of a relativistic hadronic component in blazar jets that otherwise would remain hidden. The production of narrow features or spectral hardenings due to $\pi^0$ decay in the VHE spectra of blazars is testable with the advent of CTA.
  }


   \maketitle

\section{Introduction}
Blazars are a rare class of active galactic nuclei (AGN) with relativistic jets that are powered by accretion onto a central supermassive black hole \citep{Begelman1984} and closely aligned with our line of sight \citep{Urry1995}. They are the most powerful persistent astrophysical sources of nonthermal electromagnetic radiation in the Universe, and promising candidate sources of other cosmic messengers like high-energy neutrinos \citep[for a recent review, see][]{2022arXiv220203381M}. 

The spectral energy distribution (SED) of blazars spans $\sim15$ decades in energy, starting from radio frequencies and extending to high-energy $\gamma$-rays. The flux across the electromagnetic spectrum is variable on different timescales, ranging from minutes to months, with more varying amplitudes found usually in X-rays and $\gamma$-rays~\citep[for a review, see][]{Boettcher2019}. The blazar SED has a characteristic double-humped shape, with the low-energy component peaking between infrared and X-ray energies and the high-energy component peaking in $\gamma$-rays \citep[e.g.,][]{Ulrich_1997, Fossati_1998}. The low-energy hump is usually explained by electron synchrotron radiation, whereas the origin of the high-energy component is less clear \citep[for a recent review, see][]{2020Galax...8...72C}. According to the most widely accepted scenario for high-energy blazar emission, $\gamma$-ray photons are produced via inverse Compton scattering between relativistic electrons and their own synchrotron photons \citep[synchrotron self-Compton, SSC, see e.g.,][]{1992ApJ...397L...5M, 1996ApJ...461..657B,  1997A&A...320...19M} or low-energy external radiation fields \citep[external Compton, see e.g.,][]{1992A&A...256L..27D,  1994ApJ...421..153S, 1996MNRAS.280...67G}. 

The likely association of at least some of the high-energy neutrinos detected by IceCube with AGN \citep[e.g.,][]{IceCube:2018dnn, IceCube:2018cha, Aartsen_2020,2022Sci...378..538I} has revived interest in so-called hadronic emission models. According to these, the high-energy SED component of a jetted AGN is produced directly via synchrotron radiation of protons \citep{2000NewA....5..377A, 2001APh....15..121M} or indirectly via a hadronic-initiated electromagnetic cascade \citep[e.g.,][]{1991A&A...251..723M, 1993A&A...269...67M}. Detailed SED modeling of the blazar \txs, the first astrophysical source associated with high-energy neutrinos in time and space at the $\sim3-3.5\sigma$ level~\citep{IceCube:2018dnn, IceCube:2018cha}, has revealed that the electromagnetic imprint of the proton population responsible for the neutrino signal should be subdominant, with the emission of the primary leptonic population producing the observed SED \citep{Keivani2018, 2019MNRAS.483L..12C, 2019NatAs...3...88G, 2020ApJ...891..115P}. These constraints are insensitive to the details of the models, because they basically rely on energy conservation and on the fact that electromagnetic  cascades  lead  to  broad  spectra  that extend 
in  the  X-ray  and $\gamma$-ray ranges \citep{Murase_2018}. This new class of hybrid lepto-hadronic models has been applied so far to \txs and a few other blazars that were spatially associated with high-energy neutrinos \citep{2021JCAP...10..082O,2023MNRAS.519.1396S} and they have been used to make neutrino predictions from bright \fermi-detected blazars~\citep{2019MNRAS.489.4347O, 2023arXiv230713024R}.

Recently, \citet{2020A&A...637A..86M} reported an unusual spectral feature in the very high-energy (VHE, $E>100$~GeV) spectra of the blazar Mrk~501 on the day of the highest X-ray activity  above 2 keV measured by \swift/XRT in the last 17 years. This feature at $\sim 3$~TeV was inconsistent (at a $3\sigma-4\sigma$ confidence level) with the standard analytic functions used to describe VHE spectra (e.g., a power law or a log-parabola). Instead, it was better described by a narrow Gaussian-like component or an exponential log-parabolic component. This rare feature, if true, cannot be explained by the standard one-zone SSC scenario \citep[for alternative leptonic scenarios, see][]{2020A&A...637A..86M,2021A&A...646A.115W}. This raises the question of whether there are conditions under which hadronic-related spectral features can appear at VHE during X-ray flares in the context of the usual SSC scenario for blazar emission.

In this paper, we study the emergence of narrow features in VHE $\gamma$-ray spectra of flaring blazars produced via the decay of neutral pions (henceforth, $\pi^0$ bumps).  We focus on high-synchrotron peaked (HSP) blazars~\citep[for blazar spectral classes see e.g.,][]{2010ApJ...716...30A}, as these are often detected at VHE energies \citep{2019A&A...632A..77C}. In fact, HSPs are the majority of the $\sim90$ extragalactic sources detected at VHE by imaging atmospheric Cherenkov telescopes (see TeVCat catalog\footnote{\url{tevcat.uchicago.edu}}). We perform numerical calculations of HSP SEDs during X-ray flares, reaching peak synchrotron energies $\sim 100$~keV, for various model parameters. Our goal is to search for conditions leading to the emergence of $\pi^0$ bumps in the VHE spectra of flaring blazars and apply our model to Mrk 501.

This paper is structured as follows. In Sec.~\ref{sec:model} we outline the model and the numerical code used. In Sec.~\ref{sec:results} we present the results of our numerical investigation. We then apply our model to the observations of Mrk 501 in Sec.~\ref{sec:mrk501}.  We discuss the detectability of spectral features related to $\pi^0$ decay with the Cherenkov Telescope Array (CTA) in Sec.~\ref{sec:cta}, and conclude in Sec.~\ref{sec:discussion} with a discussion of our results. We adopt a flat $\Lambda$CDM cosmology with $H_0=70$~km s$^{-1}$ Mpc$^{-1}$ and $\Omega_{\rm m}=0.3$. 

\section{The SSC+$\pi^0$ model}\label{sec:model} 
We modeled the emission region as a spherical blob of radius $R'$ containing a tangled magnetic field of strength $B'$. The blob moves with a bulk Lorentz factor, $\Gamma=\left(1-\beta_\Gamma^2 \right)^{-1/2}$, at an angle, $\theta$, with respect to the observer's line of sight. The Doppler factor was then defined as $\delta = \Gamma^{-1}(1-\beta_{\Gamma} \cos \theta)^{-1}$. Henceforth, primes will be used to denote quantities in the frame co-moving with the blob, and unprimed quantities will be used for the observer’s frame. 

We assumed that, prior to the onset of an X-ray flare, the blazar SED is fully explained by the SSC radiation of a population of relativistic electrons. These are injected continuously in the emitting region at a constant rate given by $Q_{\rm e}(\gamma'_{\rm e})=Q_{\rm e0} \gamma_{\rm e}^{'-s_{\rm e}} H(\gamma'_{\rm e};\gamma'_{\rm e,min}, \gamma'_{\rm e, max})$, where $H(x; x_1, x_2)=1$ for $x_1 \le x \le x_2$ and 0 otherwise. The normalization, $Q_{\rm e0}$, can be expressed in terms of the bolometric injection electron luminosity, $L'_{\rm e}$, as $Q_{\rm e0} = L'_{\rm e}/(m_{\rm e} c^2 F(s_{\rm e}, \gamma'_{\rm e, min}, \gamma'_{\rm e, max}))$, where $F=\ln(\gamma'_{\rm e, max}/\gamma'_{\rm e, min})$ for $s_{\rm e}=2$ and $F=(\gamma_{\rm e, max}^{'-s_{\rm e}+2}-\gamma_{\rm e, min}^{' -s_{\rm e}+2})/(2-s_{\rm e})$ otherwise. 

For hard electron distributions ($s_{\rm e}<2$), like those typically used in SSC models of HSP blazars \citep[e.g.,][]{2003ApJ...597..851K, 2015MNRAS.448..910C}, the peak synchrotron energy, $\varepsilon_{\rm s}$, is determined by the maximum electron Lorentz factor, 
\begin{equation}
 \gamma'_{\rm e, \max} \simeq  10^5 \ (1+z)^{1/2} \left(\frac{\varepsilon_{\rm s}}{3~{\rm keV}} \right)^{1/2}  \left( \frac{B'}{1~{\rm G}}\right)^{-1/2} \left(\frac{\delta}{20} \right)^{-1/2}.
 \label{eq:gemax}
\end{equation}
Therefore, the most straightforward way to simulate a flare where the synchrotron peak shifts to higher energies (henceforth, hard X-ray flares) is to assume that the maximum electron energy also increases, while the other parameters remain constant. The injection electron luminosity could be the same as or higher than its pre-flare value, but for simplicity we adopted the former option. Motivated by observations of hard X-ray flares in Mrk 501 \citep[e.g.,][]{2001ApJ...554..725T}, we assumed that the power law slope of the electron distribution does not change.

The same process that accelerates electrons can also accelerate protons to relativistic energies. To minimize the free model parameters, we assumed that the acceleration mechanism at work produces power law distributions for both particle species with  $s_{\rm p}=s_{\rm e}$ and  $\gamma'_{\rm p,\max} \sim (m_{\rm e}/m_{\rm p}) \, \gamma'_{\rm e, max}$. Similarly to electrons, protons are injected into the radiation region (blob) with a rate given by $Q_{\rm p}(\gamma'_{\rm p})=Q_{\rm p0} \gamma_{\rm p}^{'-s_{\rm p}} H(\gamma'_{\rm p};\gamma'_{\rm p,min}, \gamma'_{\rm p, max})$, where $Q_{\rm p0} = L'_{\rm p}/(m_{\rm p} c^2 F(s_{\rm p}, \gamma'_{\rm p, min}, \gamma'_{\rm p, max}))$ and $L'_{\rm p}$ is the bolometric injection proton luminosity.  

To produce TeV $\gamma$-ray photons (in the observer's frame) from $\pi^0$ decay, the parent proton energy in the comoving frame should be 
\begin{eqnarray}
\varepsilon'_{\rm p}\approx 10 \frac{\varepsilon_\gamma}{\delta}(1+z) \simeq 0.5~{\rm TeV} \ (1+z) \left(\frac{\varepsilon_{\gamma}}{1~{\rm TeV}} \right) \left(\frac{20}{\delta}\right).
\label{eq:ep}
\end{eqnarray}
In contrast to other hadronic scenarios for blazar $\gamma$-ray emission \citep[e.g.,][]{2015MNRAS.448..910C, Petropoulou_2015}, the required proton energies are low. The target photons used in  photomeson interactions with protons of energy $\varepsilon'_{\rm p}$ should have comoving energy, $\varepsilon'_{\rm t}\gtrsim \bar{\varepsilon}_{\rm th} m_{\rm p} c^2/ 2 \varepsilon'_{\rm p}$, where $\bar{\varepsilon}_{\rm th}=m_{\pi^0} c^2 (1+m_{\pi^0}/2 m_{\rm p}) \simeq 0.145$~GeV. The threshold energy condition for interactions with jet photons can then be written as 
\begin{equation}
   \left(\frac{\varepsilon_{\rm t}}{0.5~{\rm MeV}}\right) \left(\frac{\varepsilon_{\gamma}}{1~{\rm TeV}}\right) \gtrsim 11.6 \ (1+z)^{-2}  \left( \frac{\delta}{20}\right)^2\cdot
\end{equation}
Therefore, the target photons should be very energetic (hard X-rays or soft $\gamma$-rays) to enable the production of pions that will decay into TeV $\gamma$-rays (in the observer's frame).

In non-flaring states, the maximum proton energy is even lower than the one needed for the production of pionic TeV $\gamma$-rays, namely $\varepsilon'_{\rm p, max} \sim 0.05-0.15~{\rm TeV}$ for $B'\sim 0.1-1$~G and $\delta \sim 20$ (see Eqs.~\ref{eq:gemax} and \ref{eq:ep}). During hard X-ray flares with peak synchrotron energies $>100$~keV, however, the condition can be met by the most energetic protons in the blob. In this case, the bolometric luminosity of the pionic $\gamma$-rays can be estimated as $L_{\pi^0\rightarrow 2\gamma} \approx f_{p\pi} L_{\rm p}$, where  $f_{p\pi}$ is the photopion production efficiency that depends on the number density of synchrotron photons and the Doppler factor~\citep[e.g.,][]{Murase_2014}. Therefore,  $L_{\pi^0\rightarrow2\gamma}$ will be a function of $L'_{\rm e}, L'_{\rm p}, B', R'$, and $\delta$. The competition between the pionic $\gamma$-ray flux and the SSC flux in the relevant energy range, which depends on the aforementioned parameters except $L'_{\rm p}$, will determine the final shape of the VHE spectrum in the SSC+$\pi^0$ model. 

\subsection{Numerical approach}
To study the impact of the $\pi^0$ component in the VHE spectra of hard X-ray flares, we used the numerical code {\sc athe$\nu$a} \citep{1995A&A...295..613M, 2012A&A...546A.120D}, which computes the multiwavelength photon and all-flavour neutrino spectra as a function of time by solving the kinetic equations for relativistic protons, secondary electrons, and positrons, photons, neutrons, and neutrinos. The physical processes that are included in the code and couple the various particle species are: electron and proton synchrotron emission, synchrotron self-absorption, electron inverse Compton scattering, $\gamma \gamma$ pair production, photopair (Bethe-Heitler) production, and photomeson production. The latter is modeled based on the results of the Monte Carlo event
generator {\sc sophia} \citep{2000CoPhC.124..290M}, while photopair production is modeled using the Monte Carlo results of \cite{1996APh.....4..253P}; see also \cite{2005A&A...433..765M}. For the modeling of other processes we refer the reader to \cite{1995A&A...295..613M} and \cite{2012A&A...546A.120D}. With the adopted numerical scheme, energy is conserved in a self-consistent way, since all the 
energy gained by one particle species has to come from an equal amount of energy lost by another particle species. The
adopted numerical scheme is ideal for studying the development of in-source electromagnetic cascades in various regimes \citep[e.g.,][]{2014MNRAS.442.3026P, 2019ApJ...883...66P}.

We assumed that the non-flaring SED, which is described by a SSC model, has a peak synchrotron frequency, $\epsilon_{\rm s}=3$~keV, which increases to $\sim$100~keV during the flare. This was achieved by suitably increasing $\gamma'_{\rm e, max}$ and for simplicity we used a step function in time. At the same time we assumed that $\gamma'_{\rm p, max}$ undergoes a similar change, and evolved the system til it reached a new steady state characterized by higher X-ray flux and energy. A steady state was reached within $\sim 2$ light crossing times for all parameter sets we used. Our approach is sufficient for studying the flaring spectra. If, however, one would like to investigate the shape of flares at different wavelengths, a Lorentzian or Gaussian function describing the evolution of the maximum particle energy would be more appropriate \citep[e.g.,][]{2011A&A...525A..40M}.

We computed multi-wavelength flaring spectra for different values of $B'$, $R'$, and particle injection luminosities (see Table~\ref{tab:param}). These were expressed in a dimensionless form through their compactnesses, which were defined as 
\begin{equation} 
\ell_{\rm e,p} = \frac{\sigma_{\rm T} L'_{\rm e,p}}{4 \pi R' m_{\rm e, p} c^3}.
\end{equation}
In all cases, we used $s_{\rm e}=s_{\rm p}=1.7$, $\gamma'_{\rm e, min}=\gamma'_{\rm p, min}=1$, and $\delta=20$ (unless stated otherwise). For the calculation of the fluxes we considered a blazar at the redshift of Mrk~501, $z = 0.034$, which corresponds to a luminosity distance, $d_{\rm L}=149.4$~Mpc.

\begin{figure*}
    \centering
    \includegraphics[width=0.47\textwidth]{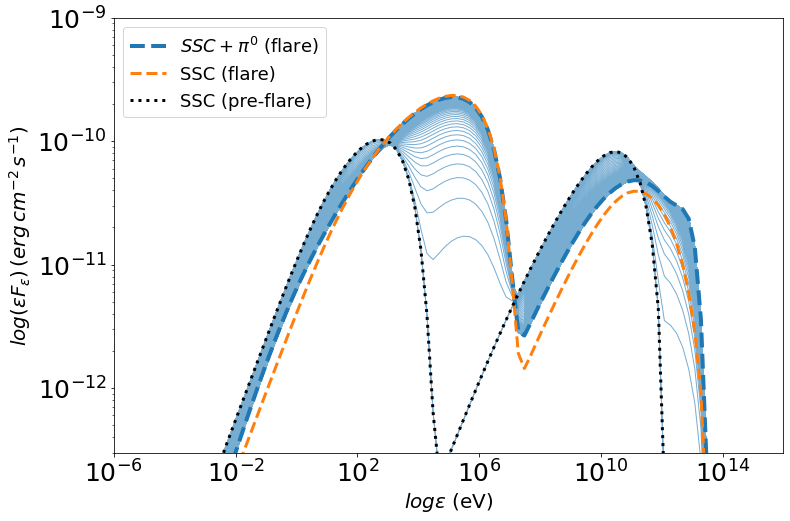}
    \hfill
    \includegraphics[width=0.47
    \textwidth]{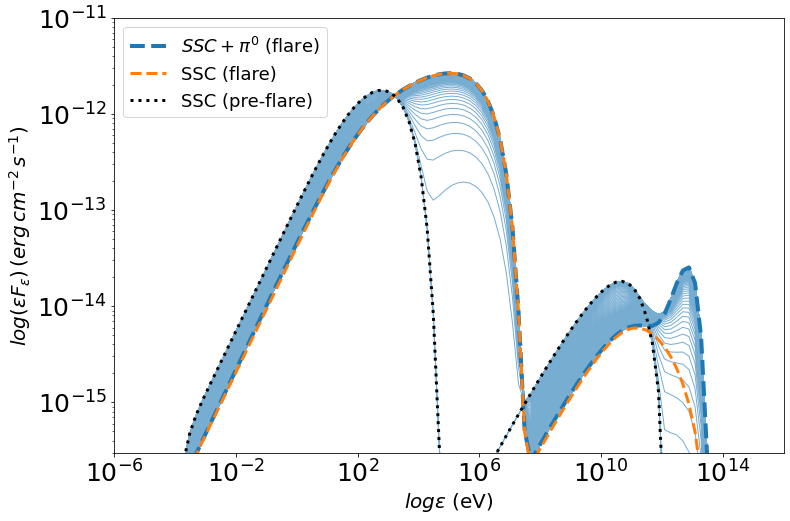}
    \caption{SED snapshots (solid lines) showing the evolution from a pre-flare steady state (dotted black line) to a new steady state that corresponds to the peak of a hard X-ray flare when protons are injected with $\gamma'_{\rm p, max}=10^{3.5}$  (dashed blue line). For comparison, the pure SSC flare spectrum is overplotted (dotted orange line). The time interval between successive snapshots is $0.1 \, t'_{\rm cr}$ and the total duration in the observer's frame is $10 \ t'_{\rm cr}\approx 4.6$~hr. The left and right panels show the results for $\ell_{\rm e}=10^{-4}$, $\ell_{\rm p}=1$ and $\ell_{\rm e}=10^{-6}$, $\ell_{\rm p}=10^{-1}$, respectively. Other parameters used are $B'=0.86$~G, $R'=10^{15}$~cm, $s_{\rm e}=s_{\rm p}=1.7$, $\gamma'_{\rm e, max}=10^5$ (pre-flare) and $10^{6.5}$ (flare), $\gamma'_{\rm p,\max} \sim (m_{\rm e}/ m_{\rm p})\gamma'_{\rm e, \max}$, and $\delta=20$. The attenuation of VHE $\gamma$-rays due to the extragalactic background light is not included here.
    }
    \label{fig1}
\end{figure*}

\section{Results}\label{sec:results}
\begin{table}
    \centering
     \caption{Input parameters for the simulation of flares with the SSC+$\pi^0$ model.}
    \begin{tabular}{ccccc}
    \hline 
     No.    & $B'$ (G) & $R'$ (cm) &  $\ell_{\rm e}$ &  $\ell_{\rm p}$ \\
     \hline
    1     & 0.86 & $10^{15}$ &  $10^{-6}$ &  $10^{-1}$ \\
    2     & 0.86 & $10^{15}$ & $10^{-5}$ & $10^{-1}$  \\
    3     &  0.86 & $10^{15}$ & $10^{-4}$  & $10^{-1}$ \\
    4     &  0.86 & $10^{15}$ & $10^{-3}$  & $10^{-1}$ \\
    5     & 0.86 & $10^{15}$ & $10^{-6}$ & $1$ \\
    6     & 0.86 & $10^{15}$ &  $10^{-5}$ &  $1$  \\
    7     &  0.86 & $10^{15}$ & $10^{-4}$ &  $1$  \\
    8     &  0.86 & $10^{15}$ & $10^{-3}$ &  $1$  \\
    9     & 0.86 & $10^{15}$ &  $10^{-6}$ &  $10$ \\
    10     & 0.86 & $10^{15}$ & $10^{-5}$ & $10$  \\
    11     &  0.86 & $10^{15}$ & $10^{-4}$ &  $10$  \\
    12     &  0.86 & $10^{15}$ & $10^{-3}$ &  $10$  \\
    \hline 
    13     &  0.86 & $10^{16}$ &  $10^{-6}$ & $10^{-1}$  \\
    14     & 0.86 & $10^{16}$ &  $10^{-5}$ & $10^{-1}$  \\
    15     &  0.86 & $10^{16}$ & $10^{-4}$   & $10^{-1}$ \\
    16     & 0.86 & $10^{16}$ & $10^{-6}$ &  $1$ \\
    17     & 0.86 & $10^{16}$ &  $10^{-5}$ & $1$  \\
    18     &  0.86 & $10^{16}$ &  $10^{-5}$ & $1$  \\
    \hline 
    19     &  86 & $10^{14}$ &  $10^{-4}$ & $10^{-0.1}$  \\  
    20     &  86 & $10^{14}$ &  $10^{-3}$ & $10^{-0.1}$  \\
    21     &  86 & $10^{14}$ &  $10^{-2}$ & $10^{-0.1}$  \\ 
    22     &  86 & $10^{14}$ & $10^{-4}$ & $1$  \\  
    23     &  86 & $10^{14}$ & $10^{-3}$ & $1$  \\      
    24     &  86 & $10^{14}$ & $10^{-2}$ & $1$  \\ 
    25     &  86 & $10^{14}$ &  $10^{-4}$ & $10$  \\ 
    26     &  86 & $10^{14}$ &  $10^{-3}$ & $10$  \\ 
    27     &  86 & $10^{14}$ &  $10^{-2}$ & $10$  \\ 
    \hline 
    \end{tabular}
    \begin{tablenotes}
     \item[] \textit{Note.} For runs 1-18 we set $\gamma'_{\rm e, max}=10^{6.5}$, while $\gamma'_{\rm e, max}=10^{5.5}$ was used for the remaining runs. In all cases, $\gamma'_{\rm p, max}=\gamma'_{\rm e, max}m_{\rm e}/m_{\rm p}$, and the particle escape rate is $t^{'-1}_{\rm esc}=c/R'$. 
     \end{tablenotes}
    \label{tab:param}
\end{table}

Figure~\ref{fig1} shows the temporal evolution of the spectrum from a pre-flare steady state (dotted black line) to a new steady state corresponding to a hard X-ray flare  (dashed blue line) for different choices of the electron and proton injection compactness. For comparison purposes, we also show the SSC spectrum of the flare (dashed orange line). In both cases, the VHE flare spectrum exceeds the SSC prediction due to the injection of energetic $\gamma$-rays from $\pi^0$ decay. The spectral signature of the latter is clearly visible on the right-hand-side panel of Fig.~\ref{fig1}, where the ratio $\ell_{\rm p}/\ell_{\rm e}$ is largest. The $\pi^0$ bump is also narrow, suggesting that only a small part of the proton energy distribution (here, the highest energy protons) satisfy the pion production threshold with the X-ray synchrotron photons. Even in a less extreme case (in terms of the proton-to-electron luminosity ratio), like the one depicted in the left-hand-side panel, there is a visible hardening of the VHE spectrum that cannot be explained by the SSC emission alone.

\begin{figure}
    \centering
    \includegraphics[width=0.47\textwidth]{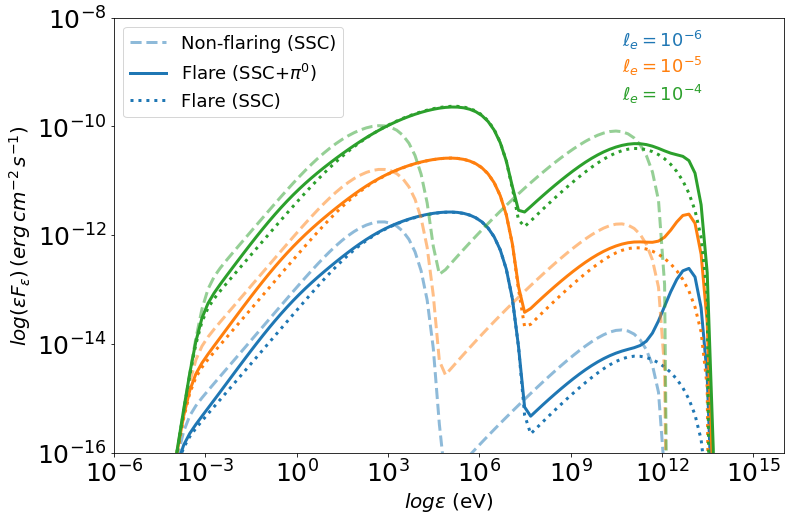}
    \caption{Non-flaring and X-ray flaring SEDs computed for three values of the electron compactness, $\ell_{\rm e}$, as is indicated in the plot and $\ell_{\rm p}=1$.  
    All other parameters are the same as in Fig.~\ref{fig1}.}
    \label{fig2}
\end{figure}

The effects of the electron injection compactness, $\ell_{\rm e}$, on the VHE spectra are better illustrated in Fig.~\ref{fig2}. Here, we plot spectra of hard X-ray flares (solid lines) computed for three values  of $\ell_{\rm e}$, while keeping all other parameters fixed. For comparison reasons, the pure SSC spectrum of the flare is also shown (dotted lines). As $\ell_{\rm e}$ increases, we recover well-known scaling relations of the synchrotron and Compton fluxes with electron luminosity \citep[e.g.,][]{1996ApJ...461..657B}. For a fixed proton luminosity, the $\pi^0$ bump becomes less prominent as the electron luminosity increases. This might seem counterintuitive at first, but it can be understood as follows. On the one hand, the number density of target photons for pion production increases almost linearly with the electron luminosity, resulting in $L_{\pi^0\rightarrow 2 \gamma} \propto \ell_{\rm e}$. On the other hand, the SSC flux scales roughly quadratically with $\ell_{\rm e}$, and thus leads to  $L_{\pi^0\rightarrow 2 \gamma}/L_{\rm SSC} \propto \ell_{\rm e}^{-1}$.

\begin{figure}
    \centering
    \includegraphics[width=0.47\textwidth]{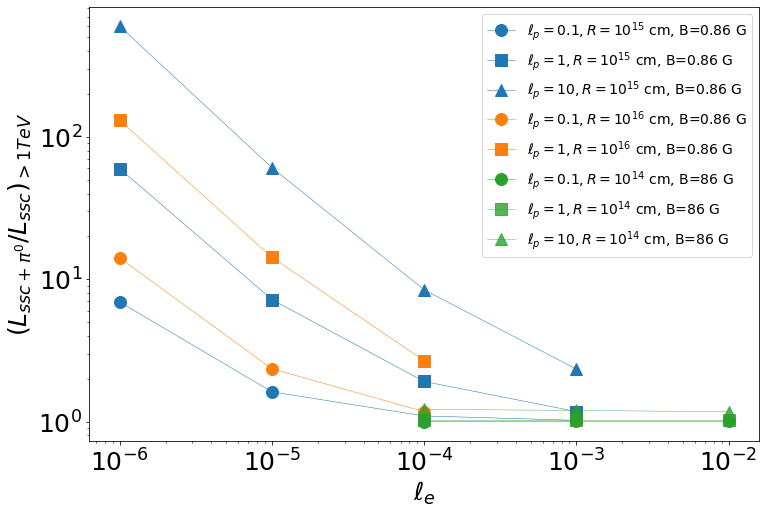}    
    \caption{Ratio of VHE luminosity ($>1$~TeV) due to SSC and $\pi^0$ decay and SSC alone as a function of the electron compactness for various parameters (see inset legend).}
    \label{fig3}
\end{figure}

The dependence of this ratio on the electron compactness is better illustrated in Fig.~\ref{fig3}, which summarizes the findings from a suite of numerical calculations performed for different parameter values (see Table~\ref{tab:param}). Here, we plot the ratio of the VHE $\gamma$-ray luminosities (integrated above 1 TeV) for the SSC+$\pi^0$ and SSC models as a function of $\ell_{\rm e}$. 
For a given proton compactness, as $\ell_{\rm e}$ increases, the VHE spectrum becomes leptonically dominated. For sufficiently high $\ell_{\rm e}$ values, the luminosity ratio should asymptotically approach unity. The contribution of the pionic $\gamma$-rays becomes evident for lower values of the electron compactness where  $\left(L_{\rm SSC+\pi^0}/L_{\rm SSC} \right)_{>1 \rm TeV} \propto \ell_{\rm e}^{-1}$. In this regime,  the luminosity ratio is proportional to $\ell_{\rm p}$, as is shown by comparing symbols of the same color. Changes in the radius of the emission region have a lesser impact on the luminosity ratio (for all other parameters fixed), as is evidenced by comparing markers of the same shape and different colors.

The findings presented in this section suggest that synchrotron-dominated flares with peak energies $\epsilon_{\rm s}\gtrsim100$~keV might be ideal periods for the search of $\pi^0-$decay bumps in the TeV spectra of HSP blazars.  In the next section we discuss the applicability of our model to the flaring spectra of Mrk 501 in 2014~\citep{2020A&A...637A..86M}.

\begin{figure*}
\centering
\includegraphics[width=0.48\textwidth]{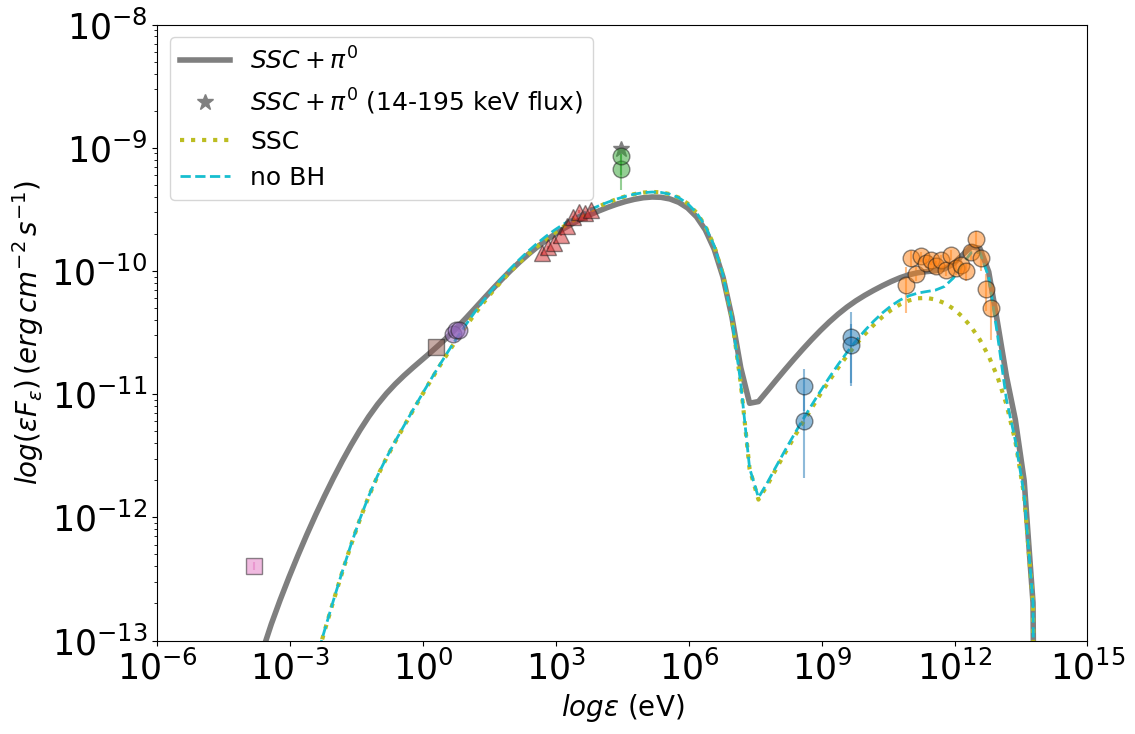}
\includegraphics[width=0.48\textwidth]{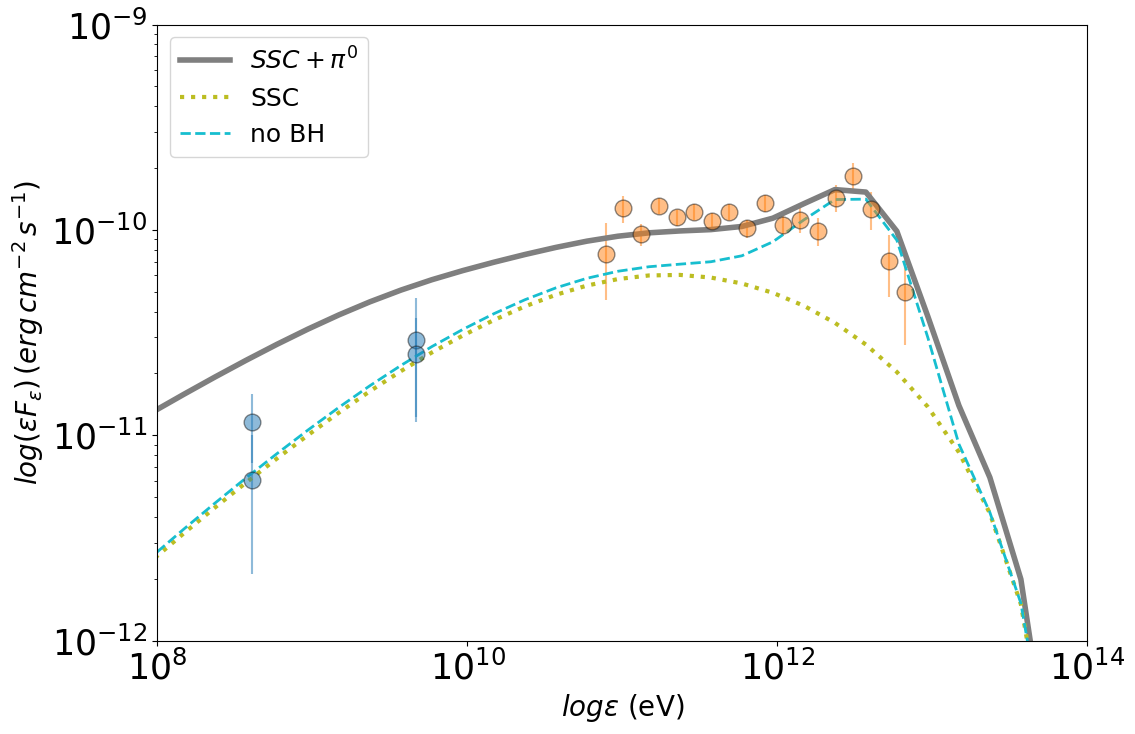}
\caption{\textit{Left panel:} Broadband SED of Mrk 501 on MJD 56857.98 compiled using data from MAGIC (orange circles, corrected for EBL attenuation), \fermi-LAT (averaged over four and ten days, blue circles), \swift-XRT (red triangles), \swift-BAT (integrated flux in 14-195 keV from one-day observations, shown with green circles), \swift-UVOT (purple circles), KVA (purple square), and  Metsähovi (pink square). The SSC+$\pi^0$ model SED is overplotted (solid line). For comparison, we also show the SED when Bethe-Heitler pair production is neglected (dashed line), and the purely leptonic (SSC) model (dotted line). \textit{Right panel:} Zoom-in of the $\gamma$-ray spectrum.} 
\label{fig:ssc-mrk501}
\end{figure*}

\section{Application to Mrk 501}\label{sec:mrk501}
Mrk 501 is a HSP blazar at $z=0.034$ that is well known for its X-ray and VHE flaring activity. It is also characterized as a transient extreme blazar. An example of such behavior was the 1997 flare, during which the peak synchrotron energy exceeded 100 keV, 
marking a change of more than two orders of magnitude in the peak energy compared to that of the non-flaring state
\citep[e.g.,][]{1998A&A...340...47P, 2001ApJ...554..725T}. However, Mrk\,501 has also shown the characteristics of an extreme blazar during non-flaring activity, as was observed throughout the extensive multi-instrument campaign in the year 2012 \citep[][]{2018A&A...620A.181A}, indicating that being an extreme blazar may not be a permanent state, but rather a state that changes over time.

\cite{2020A&A...637A..86M} reported the results of a multi-wavelength campaign performed in July 2014. For a period of about two weeks, Mrk 501 exhibited the highest X-ray activity ever detected by \swift during its  operation. VHE flaring activity correlated with the X-rays was also recorded. 
Moreover, a narrow feature at $\sim3$~TeV was detected at a significance level of about 4$\sigma$ in the VHE spectrum measured with the MAGIC telescopes on July 19, 2014 (MJD 56857.98). This was also the day with the highest X-ray flux ($>0.3$ keV) measured with \swift. We note that the spectral feature at $\sim3$~TeV may also have been present the day before and the day after (and hence may have lasted three days, from July 18 until July 20); but the detection of the spectral feature in these two days is not significant, perhaps due to the much shorter MAGIC observation time (half an hour, instead of the 1.5 hours from the observation that started on July 19). This feature is inconsistent with the empirical functions used to describe VHE spectra, namely a power law, a log-parabola, or a log-parabola with an exponential cutoff. As a result, the typical one-zone SSC model fails to explain the VHE spectra~\citep{2020A&A...637A..86M}.

Assuming that this feature was real, we applied the model presented in the previous sections to the multi-wavelength data of Mrk~501 on MJD 56857.98. Fig.~\ref{fig:ssc-mrk501} shows the respective single-night SED adopted from \cite{2020A&A...637A..86M}. Most of the data samples were selected from observations as contemporaneous as possible, taken within three hours of each other. We note however that the \fermi-LAT flux points were computed for 4-day and 10-day bins centered on  MJD~56857.98.  The reason for the longer integration time, in comparison to the X-ray and VHE gamma-ray data, is the limited sensitivity of \fermi-LAT in detecting Mrk\,501 over short timescales. Moreover, the hard X-ray flux points (green circles) represent the integrated flux in the 14–195 keV energy range that was derived using two methods \cite[for details see][]{2020A&A...637A..86M}. 

Fig.~\ref{fig:ssc-mrk501} shows a SSC+$\pi^0$ model SED of the flare (solid gray line), with the SSC contribution indicated by the dotted line (for the parameters used, see Table~\ref{tab:jet}). The VHE spectrum can be explained by the superposition of SSC emission and the $\pi^0$ bump. It should be noted that the sharpness of the spectral feature is limited by the energy resolution of the numerical code (i.e., five grid points per decade in photon energy). While for this model we used a practically monoenergetic proton distribution (see Table~\ref{tab:jet}), it is also possible to explain the bump with a hard power law proton distribution, with $s_p \sim 1.5-1.7$. In this case, however, one would need to appropriately increase the proton compactness to compensate for the ``inactive'' lower-energy protons of the distribution that would not contribute to pion production, because they would not satisfy the energy threshold for pion production with the hard X-ray photons of the flare. The fact that a monoenergetic proton distribution is not a necessary condition for the appearance of the pion bump is also supported by the numerical results presented in Fig.~\ref{fig1}, where an extended power law distribution was used. The emergence of the pion bump is instead related to the ratio of $\ell_p/\ell_e$, as is shown in Fig.~\ref{fig2}.

For a better comparison with the BAT data, we also indicate the flux of the model integrated in the same energy range (asterisk), which is consistent with the BAT value. For a more detailed spectral comparison (on a single night) between the model and the data in hard X-rays, more sensitive hard X-ray observations would be needed (e.g., observations with \textit{NuSTAR}). In addition to the $\pi^0$ bump, the synchro-Compton emission of secondary pairs from proton-photon pair production (Bethe-Heitler) is visible in the low-energy part of the spectrum (below $\sim 1$~eV) and makes a contribution to the LAT energy range too. Because of the hard ($<2$) power law slope of the primary electron distribution and the fact that the secondary emission dominates at energies $\lesssim 1$~eV, the minimum primary electron Lorentz factor is a nuisance model parameter on which only an upper limit can be placed. 

In conclusion, our results motivate an empirical spectral model that consists of two log-parabolic components or a log-parabola and an exponential cutoff power law component to describe the composite VHE spectrum, as is detailed in the next section.
 
\begin{table}
    \caption{Parameter values used in the one-zone SSC+$\pi^0$ SED models of Mrk 501 presented in Fig.~\ref{fig:ssc-mrk501}.}
    \centering
    \begin{tabular}{c c}
    \hline
     Parameter    &  Value \\
     \hline
     $\delta$    &  13 \\
     $R'$ (cm) & $1.5 \times 10^{16}$ \\
     $B'$ (cm) & $0.16$ \\
     $\ell_{\rm e}$ & $10^{-4.2}$ \\
     $s_{\rm e}$ & 1.7 \\
     $\gamma'_{\rm e, min}$ & $\le 10^4$ \\
     $\gamma'_{\rm e, max}$ & $10^{7}$ \\
     $\ell_{\rm p}$ & 1.6 \\
     $s_{\rm p}$ & 1.7 \\
     $\gamma'_{\rm p, min}$ & $10^3$ \\
     $\gamma'_{\rm p, max}$ & $10^{3.2}$ \\
     \hline
    \end{tabular}
    \label{tab:jet}
\end{table}

\section{VHE simulated spectra for CTA}\label{sec:cta}
We performed simulations of VHE spectra using the open-source software {\tt GammaPy} \citep{2017ICRC...35..766D,aguasca_cabot_arnau_2023_8033275}  to test the detectability of $\pi^0$ bumps with the CTA \citep[CTA,][]{2019scta.book.....C}. For our simulations and fitting, we followed the steps below. First, we defined the source position in the sky (using the sky coordinates of Mrk 501). Secondly, we assigned the SSC+$\pi^0$ model shown in Fig.~\ref{fig:ssc-mrk501} to the source, and derived its theoretical photon flux (in units of s$^{-1}$ cm$^{-2}$~TeV$^{-1}$), after multiplying the theoretical spectrum with $e^{-\tau_{\rm EBL}(z,E)}$, where  the optical depth, $\tau_{\rm EBL}$, was adopted from \cite{2008A&A...487..837F}. Then, we simulated the observed $\gamma$-ray spectrum using the instrument response functions (IRF) for the northern site of CTA (currently including four large-sized telescopes and nine medium-sized telescopes), for a 30-min exposure time. In all of the simulations we used a 0.11~degree extraction region centered on the source location, which for a 30-min exposure yields $\sim3,850$ source counts and $\sim$130 background counts between 0.1 TeV and 10 TeV. 

Motivated by the theoretical spectra shown in Fig.~\ref{fig:ssc-mrk501}, we used two empirical spectral models to fit the simulated data, namely a power law with an exponential cutoff (ECPL),
\begin{equation} 
\phi_{\rm ECPL}(E) = \phi_0 \cdot \left(\frac{E}{E_0}\right)^{-\Gamma} \exp(- \lambda E)
,\end{equation} 
and a dual-component model (ECPL+LP) defined as
\begin{equation}
\phi(E) = \phi_{\rm ECPL}(E) + \phi_{\rm LP}(E).
\end{equation}
The second function of the composite model is a log-parabola,
\begin{equation}
\phi_{\rm LP}(E) = \phi_0 \left( \frac{E}{E_0} \right) ^ {
  - \alpha - \beta \log{ \left( \frac{E}{E_0} \right) } },
\end{equation}
which has a peak energy (in a $E^2 \phi_{\rm LP}(E)$ versus $E$ plot) given by 
\begin{equation}
E_{\rm peak} =E_0 \ e^{(-\alpha+2)/(2\beta)}.
\label{eq:Epeak}
\end{equation}
In both models, the energies are measured in TeV units, and the flux normalization, $\phi_0$, has units of cm$^{-2}$ s$^{-1}$ TeV$^{-1}$. The pivot energy, $E_0$, in both models was frozen to 1~TeV during the fitting. 

For the spectral analysis, we fit each model to the simulated $\gamma$-ray spectrum using {\tt cash} statistics \citep{1979ApJ...228..939C}, following standardized procedures built within {\tt GammaPy}\footnote{See documentation and examples: \url{https://docs.gammapy.org}}. The fitting was performed via log-likelihood functions, so for each model a value for the fit statistics  ($-2 \log{L}$)  was returned that allowed us to compare the goodness of models.

\begin{figure*}
\centering
\includegraphics[width=0.49\textwidth]{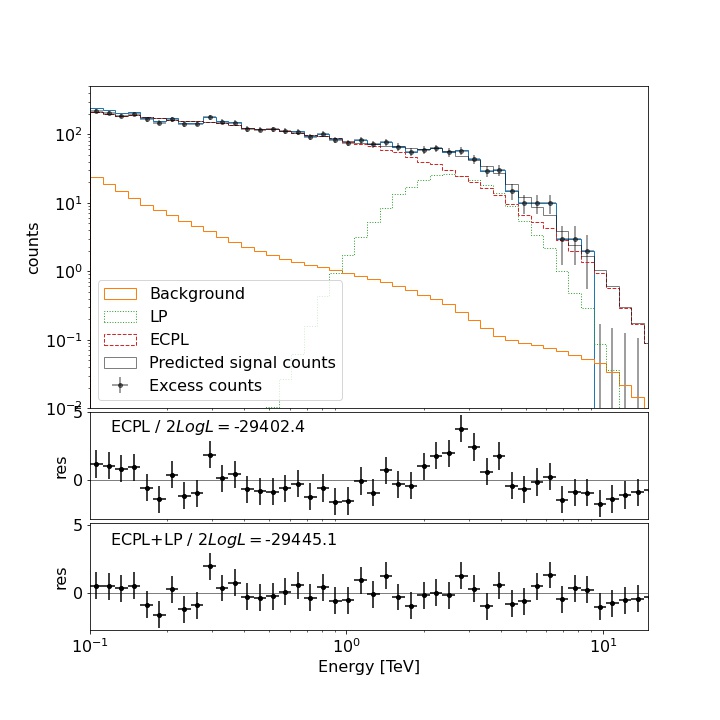}
\includegraphics[width=0.49\textwidth]{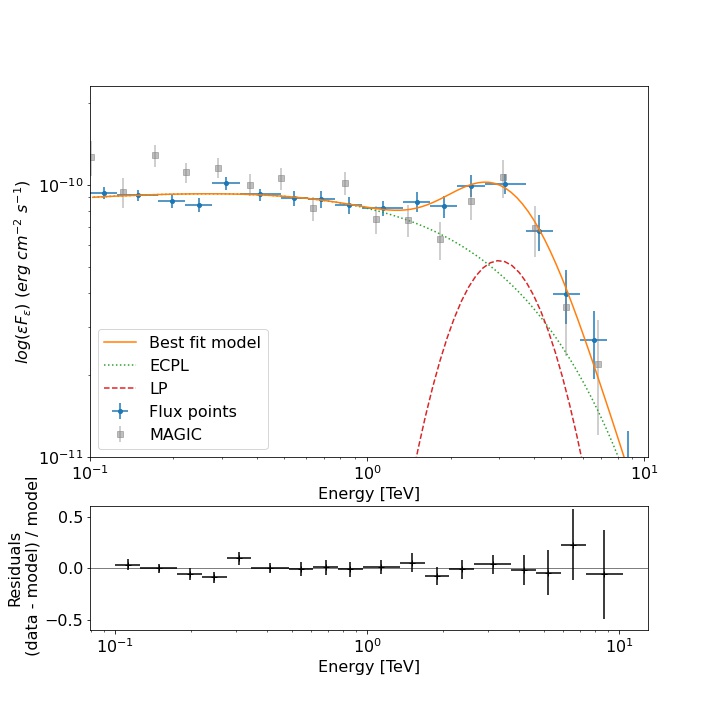}
\caption{\textit{Left panel:} Simulated $\gamma$-ray spectrum of Mrk 501 for CTA based on the SSC+$\pi^0$ model shown in Fig.~\ref{fig:ssc-mrk501} and a 30-min exposure. For illustrative purposes, the decomposition of the best-fit dual-component (ECPL+LP) model is also shown (colored histograms). The middle and bottom panels show the residuals of the fit, defined as (model-data)/error, for the ECPL and ECLP+LP models, respectively. \textit{Right panel:} Unfolded binned $\gamma$-ray spectrum (blue markers) and best-fit dual-component model (solid orange line), with the residuals shown in the bottom panel. The ECPL and LP components of the best-fit spectrum are also plotted for clarity. For comparison, the observed MAGIC spectrum (obtained with a $\sim 1.5$~hr exposure) is overplotted (gray squares).}
\label{fig:cta}
\end{figure*}

To explore the significance of any detection based on the SSC+$\pi^0$ model, we performed three tests. First, we simulated a single spectrum based on the theoretical model and fit it with the single-component (ECPL) and dual-component (ECPL+LP) models (see left panel in Fig.~\ref{fig:cta}). In addition to the improvement in the fit statistics ($\sim$43), the ECPL+LP model also improves the residuals of the fit, defined as the (model-data)/error (see middle and bottom panels of left-hand side Fig.~\ref{fig:cta}).

\begin{figure}
    \centering 
\includegraphics[width=0.45\textwidth]{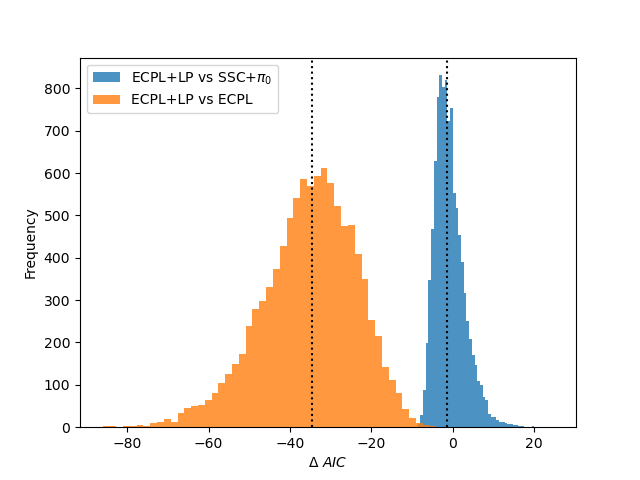}
\caption{Histograms of the difference in AIC values computed as $AIC_{Model \, 1} - AIC_{Model \, 2}$ when (i) Model 1 is the ECPL+LP model and Model 2 is the simulated model (blue), and (ii)  Model 1 is the ECPL+LP model and Model 2 is ECPL model (orange). Vertical lines indicate the median values of $\Delta  AIC$. More negative values suggest that Model 1 is more probable than Model 2. }
\label{fig:sim10K}
\end{figure}

As a second test, we repeated our simulations to establish the statistical significance of our results. We used the Akaike information criterion \citep[AIC,][]{1974ITAC...19..716A} for the model comparison. For each model we computed the AIC number by adding to the fit statistics a penalty of $2K$, where $K$ is the number of model parameters. 
To provide a statistical estimate, we repeated the procedure for 10,000 simulated spectra, and computed the difference in AIC values between two models (i.e., $\Delta  AIC = AIC_{Model \, 1} - AIC_{Model \, 2}$) to determine which model was more likely \citep[see discussion by][]{doi:10.1177/0049124104268644}.
We found that for a 30-min exposure with CTA the computed SSC+$\pi^0$ model has similar AIC values to the ECPL+LP model, while the ECPL model alone yields a worse fit, with a median $\Delta AIC\sim-34$ (see Fig. \ref{fig:sim10K}). This indicates that the ECPL+LP is preferred\footnote{The likelihood of a model over another can be estimated as $\exp(-\Delta AIC/2)$.} by more than 5$\sigma$ (based on the median $\Delta AIC$ values). Moreover, 99.8\% of simulated spectra favor the ECPL+LP model with a significance of more than 99\%. Only for a small number of simulated spectra (21 out of 10,000) do we find that $\Delta AIC > -9$ (see orange-colored histogram). We also repeated the simulations for 15-min and 60-min exposures and found that the $\Delta AIC$ values follow an almost linear trend with the exposure time. In particular, the dual-component model is favored, with a median $\Delta AIC$ of $\sim$16 and $\sim$66 over the ECPL model for 15-min and 60-min exposures, respectively. Moreover, considering the $\Delta AIC$ distribution for the 60-min exposure simulations, the $\pi^0$ bump would be detected at a median significance of 7.5$\sigma$.

 \begin{figure*}
    \centering 
\includegraphics[width=0.99\textwidth]{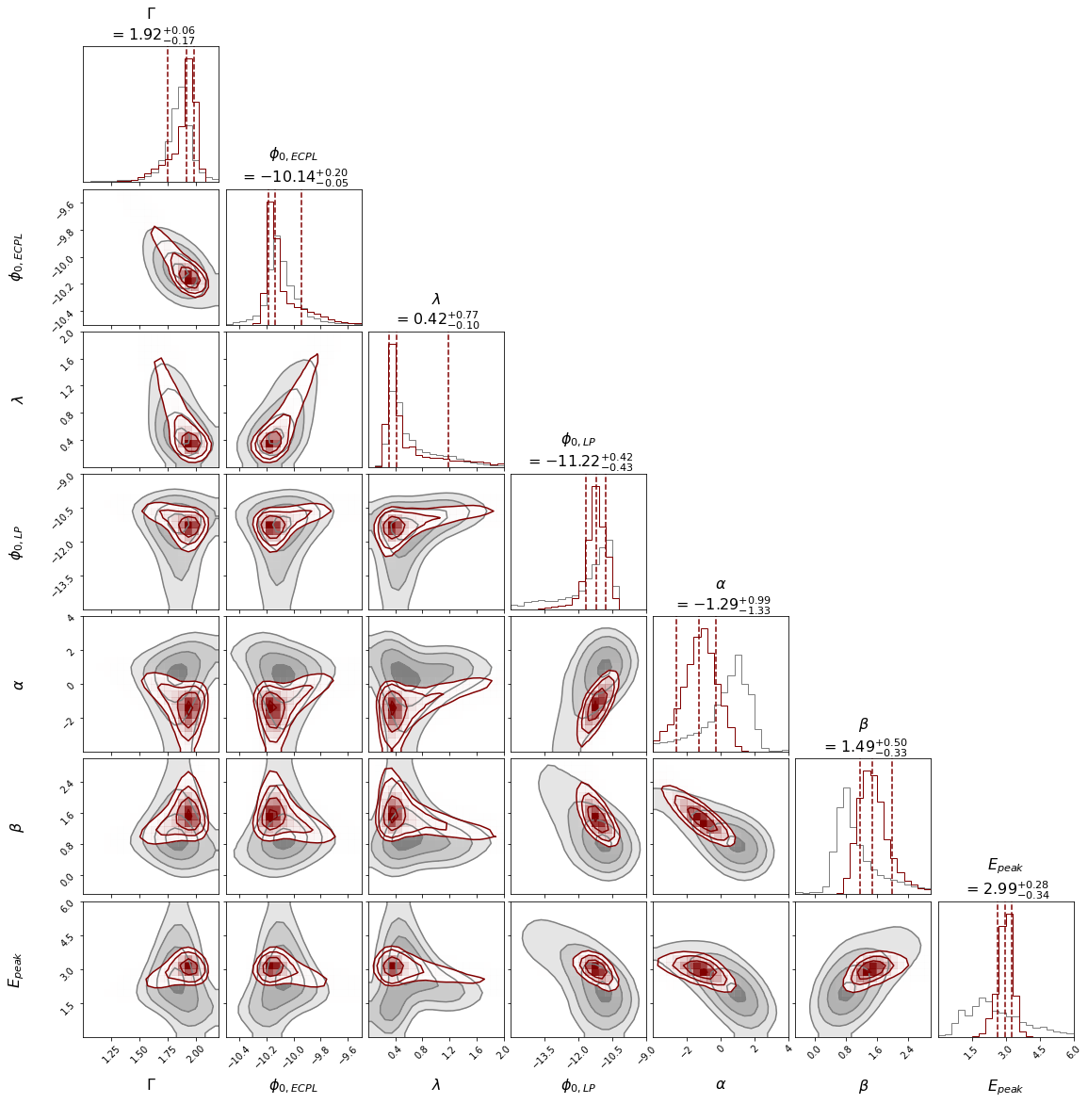}
\caption{Corner plot of the parameters obtained from the fitted models to the simulated CTA spectrum. Maroon-colored contours and distributions are based on simulations with the SSC+$\pi^0$ model fitted with the dual-component (ECPL+LP) model. The gray samples correspond to simulations based on the featureless ECPL model (i.e., no ``bump'') that are also fitted with the dual-component model. 
Contour lines indicate the 11.8\%, 39.3\%, 67.5\%, and 86.4\% levels. The parameter $E_{\rm peak}$ denotes the peak energy of the LP component (in TeV). This is not a free model parameter, but was obtained from $\alpha$ and $\beta$ -- see Eq.~(\ref{eq:Epeak}). }
\label{fig:corner}
\end{figure*}

As a final test we simulated 10,000 spectra based on a featureless broadband spectrum, using the best fit parameters of the ECPL model from the previous test. We then fit all simulated spectra with the ECPL model or the ECPL+LP model. As was expected, the $\Delta AIC$ values between these two models are clustered around zero, with a standard deviation of less than three, while less than 1\% of samples have $|\Delta AIC|>1$. This indicates that there is no statistical preference for the more complex spectral model.

The log-parabola of the dual-component model in this case does not typically describe a narrow feature, but is rather a broad component that contributes to the continuum spectrum. This is better illustrated in Fig.~\ref{fig:corner}, where we show the fitted parameters of the ECPL+LP model on the spectra that were simulated with the featureless ECPL model (light gray color). For comparison, we overplot the parameters obtained from fitting the spectra that were simulated based on the SSC+$\pi^0$ model (maroon color). We find that the ECPL parameters are consistent between the two sets of simulations. However, the LP parameters derived from the fit to the SSC+$\pi^0$ simulations are much more clustered around a central value, corresponding to a peak energy of $\simeq 3$ TeV, which is consistent with the SSC+$\pi^0$ model. On the contrary, the gray-colored contours are more dispersed, as they do not represent a sharp feature like the proposed pion bump but are a result of over-fitting the data with two components.

\section{Summary and discussion}\label{sec:discussion}
In this paper we have investigated the production of narrow spectral features in the VHE spectra of flaring blazars through $\pi^0$ decay.  During X-ray flares reaching peak synchrotron energies of $\sim 100$~keV, the number density of hard X-ray photons increases, thus making pion-production feasible for protons of a few TeV in energy. What ultimately determines the appearance of narrow $\pi^0$ bumps in the VHE spectra is the competition between the $\pi^0$ $\gamma$-ray flux and the SSC flux in the relevant energy range. We have shown that flares in HSP blazars with peak energies $\gtrsim100$~keV and small values of the Compton ratio can be ideal periods for the emergence of $\pi^0$ bumps at TeV energies.

We applied the SSC+$\pi^0$ model to multi-wavelength data of Mrk 501 that were obtained during a period of high X-ray activity that lasted two weeks. Within this \mbox{14-day} period, a narrow spectral feature at $\sim 3$~TeV was detected with MAGIC on MJD 56857.98 at a $\sim 3\sigma-4\sigma$ confidence level, and it could have been present for a maximum of three consecutive nights~\citep{2020A&A...637A..86M}. Here, we focused on single-night data of MJD 56857.98, and searched for a steady-state SSC+$\pi^0$ model that describes the broadband spectrum well. We showed that the VHE $\gamma$-ray spectrum can be explained by a superposition of $\gamma$-ray emission by neutral pion decay, inverse Compton scattering emission of secondary leptons produced via proton-photon pair production, and SSC emission of primary electrons. For a good description of the narrow spectral feature, the proton distribution has to be energetically dominated by protons with energy $E_{p} \sim 10~{\rm TeV} \, (\epsilon_\gamma/1~\rm TeV)$. This could be realized by either a mono-energetic proton distribution or by a hard ($s_p < 2$) power law distribution extending to lower energies, as was discussed in Sec.~\ref{sec:mrk501}. 

Given that the effective photopion production cross section is $\sim 10^{-3} \sigma_T$ and that the number density of the X-ray target photons is low, a very high proton luminosity is needed to produce a pionic bump visible in the TeV spectrum. Let us consider, for example, protons with Lorentz factor $\gamma^\prime_p \sim 10^{3.2}$.
These protons produce pions at a threshold with photons of comoving energy 
$\sim 90$~keV, whose number density is $n^\prime_X \simeq 10^3$~cm$^{-3}$ for the model spectrum shown in Fig.~\ref{fig:ssc-mrk501}. The photopion production efficiency can be estimated as $f_{p\gamma} = t'_{esc}/t'_{p\gamma} \approx R^\prime n^\prime_X \hat{\sigma}_{p\gamma}  \simeq 10^{-9}$, where $t_{p\gamma}$ is the proton energy loss timescale due to the photopion production process and $\hat{\sigma}_{p\gamma}  \simeq 7 \cdot 10^{-29}~\rm cm^2$ is the cross section of the process weighted by the inelasticity~\citep{2009HEA_BH}. The amount of energy transferred from relativistic protons to secondaries per unit time (in the observer's frame) can then be estimated as $L_{p\gamma} \approx f_{p\gamma} \delta^4 L^\prime_{p} \simeq 10^{44}~{\rm erg s^{-1}}\ \delta_1^4 L^\prime_{p,49}$, where we introduced the notation $q_m = 10^m$ (in cgs units). Therefore, very high proton luminosities are required to compensate for the extremely low pion-production efficiency in the SSC+$\pi^0$ model. 

Furthermore, the flaring region that produces the TeV pion bump is particle dominated, with $u^\prime_p \gg u^\prime_e \gtrsim u^\prime_B$. For the parameter values listed in Table~\ref{tab:jet}, the relativistic proton number density is $n^\prime_{p} \sim u^\prime_{p}/(\gamma^\prime_p m_p c^2) \sim  \ell_{\rm p} / (3 \sigma_T R^\prime \gamma^\prime_p) \simeq  3\times 10^4$~cm$^{-3}$ for $\gamma^\prime_p \sim 10^{3.2}$. The total number density in relativistic protons would be even higher if, instead of an effectively mono-energetic proton distribution, an extended power law starting from $\gamma'_{p, \min}=1$ were adopted. For instance, we find similar VHE gamma-ray spectra to those displayed in Fig.~\ref{fig:ssc-mrk501} for $n'_{p, \rm PL}/n'_p \sim 166$, where the subscript ``PL'' indicates an extended power law distribution with index $s_p=1.5$. The average jet power may be determined by the conditions during non-flaring states, assuming that pionic bumps, like the feature identified in Mrk~501, are rare. If almost all cold protons in the jet region producing the flare become relativistic with $\gamma'_p \sim 10^3$ during the flare, we can set a lower limit to the kinetic power of the jet in non-flaring states, which is attributed to the cold protons, namely $L_{p,j}^{(c)} > \pi R^2 \Gamma_j^2 \beta_j c n^\prime_{p, c} m_p c^2 \simeq 1.8 \times 10^{47}$~erg s$^{-1}$ for $n'_{p,\rm c} = n'_p$ (or $3.6\times10^{49}$ erg s$^{-1}$ for  $n'_{p,\rm c} = n'_{p, \rm PL}$),  $\Gamma_j \sim \delta = 13$, and $\beta_j = 1$. The lower limit on the jet power estimated for non-flaring conditions can readily exceed the Eddington luminosity of the central black hole in Mrk~501, for a typical mass of $M_{\rm BH}=10^9~M_{\odot}$ \citep[inferred by the $M_{\rm BH}-\sigma_*$ relation,][]{Barth_2002}. Therefore, the SSC+$\pi^0$ model for TeV flares requires a proton-dominated flaring region, protons reaching energies not much higher than a few tens of TeV, and super-Eddington average jet powers. These features, except for the lower maximal proton energies, are similar to those found in other leptohadronic models for blazar emission \citep[e.g.,][]{Petropoulou_2015,2015MNRAS.450L..21Z}.

In \cite{Gamma2022_Petropoulou} we considered a different scenario where relativistic protons interact with an external source of X-ray radiation (e.g., from a radiatively inefficient accretion disk), which is not directly observed, to produce the pion bump. In this case, the regions producing the X-ray flare observed with \swift and the VHE emission are decoupled and are described by different physical parameters. In particular, the ``SSC zone'' has a weaker magnetic field and is more extended than the ``pion zone,'' which is assumed to be located closer to the base of the jet \citep[for details, see Table 1 in][]{Gamma2022_Petropoulou}. 
The former zone can fully account for the optical-X-ray emission of the flare and produce a fraction of the $\gamma$-ray
flux, while the ``pion zone'' can dominate the emission at GeV-TeV energies during the flare (see Fig.~2 in \cite{Gamma2022_Petropoulou}).
In this scenario, the appearance of the $\pi^0$ bump during a hard X-ray flare, as was observed in Mrk~501, should be coincidental. Flares from the SSC zone or pion zone would also result in different strengths of correlation between X-ray and TeV $\gamma$-ray fluxes. However, this model would require the presence of a persistent ``pion zone'' to explain a weak X-ray-TeV correlation during non-flaring epochs as was observed, which would also imply a very high (time-average) jet power. 

A distinct spectral feature like the $\pi^0$ bump could be detected by future CTA observations, as we demonstrated in Sec.~\ref{sec:cta}. In particular, we performed simulations based on the theoretical SSC+$\pi^0$ model shown in Fig.~\ref{fig:ssc-mrk501} using the ``Alpha Configuration'' of the CTA Northern Array. We showed that for events collected within a 30-min CTA observation of Mrk 501, spectral fitting would require a multi-component empirical model, and a feature like the 3~TeV bump could be detected at a 5$\sigma$ significance level.

Our analysis opens up some interesting possibilities for the interpretation of multi-wavelength blazar spectra during flaring states. We demonstrated that a harder VHE $\gamma$-ray spectrum than the usual SSC component or, more occasionally, a distinct neutral pion bump at VHE energies during hard X-ray flares, can be suggestive of a relativistic hadronic component in the jet that otherwise would remain hidden. Further testing of this possibility requires simultaneous and detailed spectral observations in the soft and hard X-rays and TeV $\gamma-$ rays that can be performed by the current generation of X-ray instruments (e.g., \textit{NuSTAR}) and Cherenkov Telescopes (namely, H.E.S.S., MAGIC, and VERITAS) or, even better, by more sensitive future instruments such as CTA and X-ray satellites with broadband capabilities, like HEX-P \citep{Madsen2019}.

\begin{acknowledgements}
The authors would like to thank the anonymous referee for a constructive report. MP acknowledges support from the MERAC Fondation through the project THRILL and from the Hellenic Foundation for Research and Innovation (H.F.R.I.) under the ``2nd call for H.F.R.I. Research Projects to support Faculty members and Researchers'' through the project UNTRAPHOB (Project ID 3013). GV acknowledges support by Hellenic Foundation for Research and Innovation (H.F.R.I.) under the ``3rd Call for H.F.R.I. Research Projects to support Postdoctoral Researchers'' through the project ASTRAPE (Project ID 7802). JBG acknowledges financial support from the Spanish Ministry of Science and Innovation (MICINN) through the Spanish State Research Agency, under Severo Ochoa Program 2020-2023 (CEX2019-000920-S). D.P. acknowledges support from the Deutsche Forschungsgemeinschaft (DFG; German Research Foundation) under Germany's Excellence Strategy EXC-2094—390783311.

\end{acknowledgements}
  
  \bibliographystyle{aa} 
  \bibliography{references} 
\end{document}